\documentclass[12pt]{article}
\usepackage{graphics}
\newcommand{\rlpartial}{\partial^{{}^{{}^{\!\!\!\!\!\!\!\longleftrightarrow}}}}

\renewcommand{\theequation}{\arabic{section}.\arabic{equation}}
\begin{document}
\markright{Nonthermal nature of...}
\title{Nonthermal nature of incipient extremal black holes}
\author{Stefano Liberati~$^*$, Tony Rothman~$^\dagger$ and
Sebastiano Sonego~$^\ddagger$\\[2mm] {\small \it
\thanks{liberati@sissa.it}~
International School for Advanced Studies, via Beirut 2-4, Trieste
34014, Italy;}\\ {\small \it INFN sezione di Trieste.}\\ {\small
\it
\thanks{trothman@titan.iwu.edu}~Dept.\ of Physics, Illinois
Wesleyan University, Bloomington, IL 61702, USA.}\\ {\small
\it
\thanks{sebastiano.sonego@dic.uniud.it}~Universit\`a di
Udine, Via delle Scienze 208, 33100 Udine, Italy.}\\ }
\date{{\small 17 March 2000; \LaTeX-ed \today}}
\maketitle
\begin{abstract}
  We examine particle production from spherical bodies collapsing into
  extremal Reissner-Nordstr\"om black holes.  Kruskal coordinates
  become ill-defined in the extremal case, but we are able to find a
  simple generalization of them that is good in this limit. The
  extension allows us to calculate the late-time worldline of the
  center of the collapsing star, thus establishing a correspondence
  with a uniformly accelerated mirror in Minkowski spacetime. The
  spectrum of created particles associated with such uniform
  acceleration is nonthermal, indicating that a temperature is not
  defined.  Moreover, the spectrum contains a constant that depends on
  the history of the collapsing object. At first sight this points to
  a violation of the no-hair theorems; however, the expectation value
  of the stress-energy-momentum tensor is zero and its variance
  vanishes as a power law at late times.  Hence, both the no-hair
  theorems and the cosmic censorship conjecture are preserved. The
  power-law decay of the variance is in distinction to the exponential
  fall-off of a nonextremal black hole. Therefore, although the
  vanishing of the stress tensor's expectation value is consistent
  with a thermal state at zero temperature, the incipient black hole
  does not behave as a thermal object at any time and cannot be
  regarded as the thermodynamic limit of a nonextremal black hole,
  regardless of the fact that the final product of collapse is
  quiescent.

\vspace*{5mm} \noindent PACS: 04.62.+v, 04.70.Dy, 04.20.Dw
\end{abstract}

\section{Introduction}
\label{sec1}

Extremal black hole solutions have long played a prominent role in
black-hole thermodynamics. Early on, investigators realized that
the zero surface gravity of extremal black holes, which implies
zero Hawking temperature, makes them the natural equivalent of the
zero temperature states in ordinary thermodynamics.

Nevertheless, the third law of black-hole dynamics
\cite{BCH73,Israel86} states that the zero temperature state (the
extremal black hole) is unattainable by means of a finite number
of physical processes. The real status and meaning of this law is
still subject of debate and investigation \cite{Wald97}, but
recently a point of view has emerged, according to which extremal
black holes are thermodynamically different from the zero
temperature limit of non-extremal ones
\cite{PSSTW91,GM97,KL99,Belgiorno99}.

Over the past five years, advances in string theory \cite{Strom96}
have also stimulated a resurgence of interest in extremal
solutions. These have played a prominent role in both D-brane and
supergravity calculations of black-hole entropy and these results
seem to imply, contrary to what might be inferred from above, that
the Bekenstein-Hawking relationship between entropy and area holds
in the extremal case.

Semi-classical calculations, on the other hand, have thus far
corroborated the conclusion implied by the third law, that the
nature of extremal black holes intrinsically differs from that of
nonextremal ones. In particular, such calculations predict a
vanishing entropy for extremal black holes
\cite{KL99,GK95,HHR95,HH96,LP97}, contradicting the string-theory
results.

Given the apparent incompatibility between the two approaches, and
the fact that it might indicate some nontrivial issue in the
low-energy limit of superstring theories, we try here to improve
our understanding of the nature of extremal black holes from a
semiclassical point of view. However, we shall not deal with the
interpretation of the high-energy results in the present work,
leaving this issue for future investigations.

The calculations cited above have mainly dealt with eternal black
holes. It is thus unclear whether the thermodynamic discontinuity just
mentioned applies to the case of black holes formed by collapse. For
this reason we have decided to examine particle production by an
``incipient'' Reissner--Nordstr\"om (RN) black hole: A spherically
symmetric collapsing charged body whose exterior metric is RN. In this
paper we do not address the issue of actually constructing solutions
of the Einstein equations that describe the collapse of charged
configurations, because some simple solutions of this kind have
already been found \cite{Bou73,FH79,P83}. We emphasize that one of our
main results is that the incipient extremal black hole does radiate in
the early stages of the collapse. The fine-tuning which would then be
required to produce the extremal solutions makes the former assumption
of their existence highly nontrivial, because they would be extremely
sensitive to effects such as the backreaction of the quantum radiation
on the metric; we discuss these matters further in the conclusion.
Nevertheless, for our purposes we assume that models can be found in
which collapse leads to a black hole with $Q^2=M^2$.

We approach the problem in standard fashion, modeling the collapse
by a mirror moving in two-dimensional Minkowski spacetime
\cite{FD77}. The spectra resulting from the mirror's worldline
will then be the same as that of the black hole, up to gray-body
factors due to the nontrivial metric coefficients of RN spacetime
and to the different dimensionality. However, to determine the
appropriate worldline for the mirror one must employ coordinates
that are regular on the event horizon, and although we find the
tortoise coordinate $r_*$ to be continuous at the $Q^2=M^2$ limit,
the usual Kruskal transformation fails there. Nevertheless, we
provide a natural extension of Kruskal coordinates that is good
for the $Q^2=M^2$ case. The transformation cannot be explicitly
inverted in terms of elementary functions, but is suitable for
obtaining the asymptotic behavior for the collapsing star. This
leads us to consider a uniformly accelerated mirror in Minkowski
spacetime, whose spectrum of created particles is nonthermal. We
therefore conclude that incipient extremal RN black holes create
particles with a nonthermal spectrum.

We find, moreover, that the spectrum's amplitude contains a constant
that depends on the history of the collapsing object, apparently
violating the no-hair theorems. However, the expectation value of the
particles' stress-energy-momentum tensor is zero and its variance
vanishes as a power law at late times. Consequently, particle creation
dies out in the late stages of collapse, and is such that both the
no-hair theorem and the cosmic censorship conjecture are preserved.
One might argue that the zero value of the physical
stress-energy-momentum tensor is consistent with a thermodynamic
object at zero temperature. True enough, however, as we will see, the
{\em approach\/} to zero of the stress tensor and its variance along
with the non-Planckian spectrum indicate that the collapsing body acts
like a thermal body at no time in its history. Therefore, although the
final object is quiescent, it is improper to regard it as the zero
temperature limit of a nonextremal black hole.

\section{Kruskal-like coordinates for the extremal
RN solution} \setcounter{equation}{0} \label{sec2}

Several textbooks in general relativity (see, e.g., Refs.\
\cite{Misner73,Hawking73}) imply that Carter \cite{Carter66} found
the maximal analytical extension of RN spacetime for $Q^2=M^2$. In
fact he made a very ingenious qualitative analysis without
actually providing an analog of the Kruskal coordinates for the
extremal case. Nevertheless, for our analysis it is essential to
have such a coordinate transformation. For this reason we are
going to retrace the steps leading to the maximal analytic
extension of RN, paying close attention to the difference between
the nonextremal and extremal situations.

The first step in the procedure is to define the so-called
``tortoise'' coordinate, which is then used to construct the
Kruskal coordinates. We start with the usual form of the RN
geometry,
\begin{equation}
{\rm d}s^2 = -\left(1- \frac{2M}{r} + \frac{Q^2}{r^2}\right){\rm
d}t^2
    + \left(1- \frac{2M}{r} +
    \frac{Q^2}{r^2}\right)^{-1}{\rm
    d}r^2+r^2\, {\rm d}\Omega^2\;,
\end{equation}
where ${\rm d}\Omega^2$ is the metric on the unit sphere. The
tortoise coordinate $r_*(Q,M)$ is given by
\begin{equation}
r_*(Q,M) = \int \frac{{\rm d}r}{\left(1-2M/r+Q^2/r^2\right)}\;.
\label{tort(QM)}
\end{equation}
Carrying out the integration yields, for the nonextremal case,
\begin{equation}
r_*(Q,M) = r + \frac{1}{2\sqrt{M^2 - Q^2}}
       \left(r_{+}^2\ln(r-r_{+})-r_{-}^2 \ln(r -
       r_{-})\right) + \mbox{const},
\label{inttort(QM)}
\end{equation}
where as usual $r_{\pm}= M \pm \sqrt{M^2-Q^2}$.

Now, if we set $Q^2 = M^2$ in Eq.\ (\ref{tort(QM)}) {\it before\/}
integrating, we find the ``extremal'' $r_*$:
\begin{equation}
r_*(M,M) = r + 2M\left(\ln(r-M) - \frac{M}{2(r-M)}\right)
        + \mbox{const}.
\label{inttort(MM)}
\end{equation}
Note that the coordinate $r_*(M,M)$ diverges only at $r = M$, but
setting $Q^2= M^2$ in $r_*(Q,M)$ appears to yield the
indeterminate form $0/0$. However, if we let $Q^2 =
M^2(1-\epsilon^2)$, with $\epsilon \ll 1$, and work to first order
in $\epsilon$, it is straightforward to show that Eq.\
(\ref{inttort(QM)}) does reduce to Eq.\ (\ref{inttort(MM)}).
Therefore $r_*$ is continuous even at extremality.

Unfortunately, the Kruskal transformation itself breaks down at
that point. The Kruskal transformation is
\begin{equation}
\left.
\begin{array}{lll} {\cal U} = -{\rm e}^{-\kappa u}
& \Leftrightarrow & {\displaystyle u= -{1\over\kappa}\ln({\cal
-U})}\\ {\cal V} = {\rm e}^{\kappa v} & \Leftrightarrow &
{\displaystyle v = {1\over\kappa}\ln {\cal V}}
\end{array} \right\}\;,
\label{Kruskal}
\end{equation}
where
\begin{equation}
\left.
\begin{array}{l}
u = t - r_*\\ v = t+r_*
\end{array}
\right\} \label{EF}
\end{equation}
are the retarded and advanced Eddington-Finkelstein coordinates,
respectively, and $\kappa$ is the surface gravity. The latter is
defined as
\begin{equation}
\kappa = \lim_{r\to r_+}{1\over 2}{{\rm d} \over {\rm d}r}
\left(1-{2M\over r} + {Q^2\over r^2}\right)= {\sqrt{M^2-Q^2}\over
r_+^2}\;,
\end{equation}
and vanishes for $Q^2 = M^2$. Therefore the Kruskal coordinates
$\cal U$ and $\cal V$ become constant for any value of $u$ and $v$
and so the transformation (\ref{Kruskal}) becomes ill-defined at
that point.

We are nonetheless able to remedy this situation. Note that the
Eddington-Finkelstein coordinates are constructed by adding or
subtracting $r_*$ to $t$, as in Eqs.\ (\ref{EF}) above. Now, for
the extremal case, $r_*$ is given by Eq.\ (\ref{inttort(MM)}),
which has the extra pole $M^2/(r-M)$ with respect to the strictly
logarithmic dependence of the Schwarzschild and nonextremal RN
cases (compare Eq. (\ref{inttort(QM)})). The simplest thing to do
is define a function
\begin{equation}
\psi(\xi)= 4M\left(\ln\xi - \frac{M}{2\xi}\right) \label{psi}
\end{equation}
and guess that a suitable generalization of the Kruskal
transformation is
\begin{equation}
\left.
\begin{array}{l}
u = -\psi(-{\cal U})\\ v = \psi({\cal V})
\end{array}\right\}.
\label{Gen}
\end{equation}
Note that $\psi'(\xi)=4M/\xi+2M^2/\xi^2
> 0$, always, and so $\psi$ is monotonic; therefore
(\ref{Gen}) is a well-defined coordinate transformation. Note also
that
\begin{equation}
r_*(M,M) = r+\frac{1}{2}\psi (r-M) ,
\end{equation}
which means that near the horizon\footnote{Hereafter, for two
functions $f$ and $g$, we use the notation $f\sim g$ to mean $\lim
f/g=1$ in some asymptotic regime.}
\begin{equation}
r_*(M,M) \sim {1\over 2}\,\psi (r-M)\;. \label{r*(r-M)}
\end{equation}

We can give our choice of $\psi$ added motivation by noting that
near the horizon Eq.\ (\ref{inttort(QM)}) gives
\begin{equation}
r_*(Q,M)\sim {1\over 2\kappa}\,\ln(r-r_+)\;. \label{boia}
\end{equation}
Thus we see that the function $\kappa^{-1}\ln(\cdots)$ that enters
in the transformation (\ref{Kruskal}) from the Kruskal to the
Eddington-Finkelstein coordinates, is just twice the one which
gives a singular contribution to $r_\ast(Q,M)$ at $r=r_+$. Our
extension (\ref{Gen}) is therefore analogous to the Kruskal
transformation (\ref{Kruskal}): We choose $\psi$ as the part of
$r_\ast$ that is singular at $r=r_+$, a procedure that should work
in other, similar situations.

For (\ref{Gen}) to be a good coordinate extension, the new
coordinates $\cal U$ and $\cal V$ must be regular on the event
horizon, ${\cal H}$. This will be the case if the metric after the
coordinate transformation is singular only at $r = 0$. For the
extremal case the metric in terms of $u$ and $v$ reads
\begin{equation}
{\rm d}s^2 = -\left(1 - {M\over r}\right)^2 {\rm d}u\, {\rm d}v +
r^2\,{\rm d}\Omega^2\;. \label{Krmetric}
\end{equation}
Written in terms of {the ``Kruskal-like'' coordinates} $\cal U$
and $\cal V$, the metric (\ref{Krmetric}) assumes the form
\begin{equation}
{\rm d}s^2 = -\frac{(r-M)^2}{r^2} \psi'(-{\cal U})\psi'({\cal
V}){\rm d}{\cal U}{\rm d}{\cal V}
    +r^2\,{\rm d}\Omega^2\;.
\label{Genmetric}
\end{equation}
This line element apparently is degenerate at ${\cal H}$; if so
the transformation is ill-defined there. We now show, however,
that the factor $(r-M)^2$ is actually killed and that the
transformation is regular at $r = M$.

At ${\cal H}$ the coordinate $v$ is always finite and so
asymptotically we have $t\sim -r_*$. Therefore $u\sim -2r_*\sim
-\psi(r-M)$ where the last approximation follows from Eq.\
(\ref{r*(r-M)}). The inverse transformation yields
\begin{equation}
{\cal U}=-\psi^{-1}\left(-u\right)\sim-\psi^{-1}
\left(\psi(r-M)\right)=-\left(r-M\right)\;.
\end{equation}
Then, from the expression for $\psi'$ given above we have near the
horizon
\begin{equation}
\psi'(-{\cal U})\sim \frac{4M}{r-M}+\frac{2M^2}{(r-M)^2}\sim
\frac{2M^2}{(r-M)^{2}}\;. \label{psi'}
\end{equation}
Furthermore, since ${\cal V}$ is everywhere nonzero and finite
then $\psi'({\cal V})$ is regular there. Now it is easy to see
that the form taken by the metric (\ref{Genmetric}) is
asymptotically
\begin{equation}
{\rm d}s^{2}\sim -\frac{2M^2}{r^{2}}\psi'({\cal V}) {\rm d}{\cal
U}{\rm d}{\cal V}+r^2\,{\rm d}\Omega^2\;. \label{Finmetric}
\end{equation}
The $(r-M)^2$ in the numerator of Eq.\ (\ref{Genmetric}) is killed
by the $(r-M)^2$ in the denominator of Eq.\ (\ref{psi'}).
Consequently, ${\cal U}$ and ${\cal V}$ are good Kruskal-like
coordinates.

Notice that the coordinates $u$ and $v$ defined by the
transformation (\ref{Kruskal}) do not tend to those given by
(\ref{Gen}) as $Q^2\to M^2$. This is related to the fact that the
maximal analytic extensions of RN spacetime are qualitatively
different in the two cases \cite{Hawking73}, and is another
evidence of the discontinuous behaviour mentioned in the
Introduction.

\section{Asymptotic worldlines}
\setcounter{equation}{0}
 \label{sec3}

With the result of the previous section in hand we are now able to
construct late-time asymptotic solutions for the incipient
extremal black hole. Our goal is to find an equation for the
center of the collapsing star (in the coordinates $u$ and $v$)
that is valid at late times. Equation (\ref{EF}) gives $u$ and $v$
outside the collapsing star, but the center of the star, of
course, is in the interior. We must therefore extend $u$ and $v$
to the interior. Since $u$ and $v$ are null coordinates,
representing out- and in-going light rays, respectively, the
extension can be accomplished almost trivially by associating any
event on the interior of the star with the $u$ and $v$ values of
the light rays that intersect at this event.

The most general form of the metric for the interior of a
spherically symmetric star can be written as
\begin{equation}
{\rm d}s^2 = \gamma(\tau,\chi)^2(-{\rm d}\tau^2 + {\rm d}\chi^2) +
\rho(\tau,\chi)^2\,{\rm d}\Omega^2\;,
\end{equation}
where $\gamma$ and $\rho$ are functions that can be chosen to be
regular on the horizon. From the coordinates $\tau$ and $\chi$ we
can construct interior null coordinates $U = \tau - \chi$ and $V =
\tau + \chi$, which will also be regular on the horizon. The
center of the star can be taken at $\chi = 0$, in which case $V =
U$ and ${\rm d}V = {\rm d}U$ there (see Fig.\ 1).
%
\begin{figure}[hbt]
\vbox{ \hfil \scalebox{0.40}{{\includegraphics{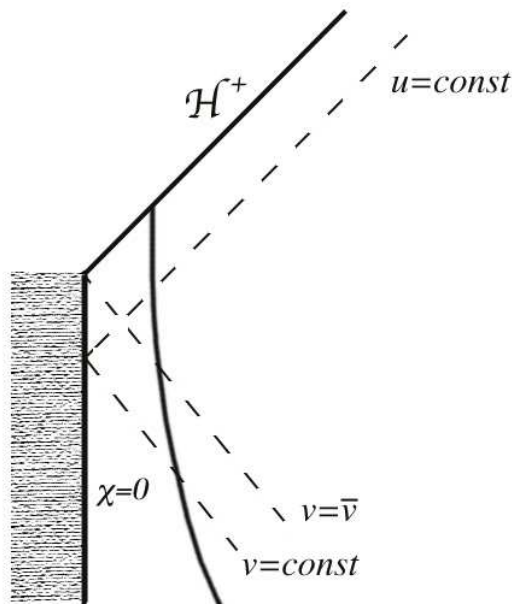}}} \hfil}
\bigskip
\caption{
A representation of gravitational collapse in null coordinates.
The portion of spacetime beyond the event horizon ${\cal H}^+$ is
not shown.
} \label{star}
\end{figure}
%
Because the Kruskal coordinates $\cal U$ and $\cal V$ are regular
everywhere as well, they can be matched to $U$ and $V$. In
particular, if two nearby outgoing rays differ by ${\rm d}U$
inside the star, then they will also differ by ${\rm
d}U=\beta({\cal U}) {\rm d}{\cal U}$, with $\beta$ a regular
function, outside. By the same token, since $V$ and $v$ are
regular everywhere, we have ${\rm d}V = \zeta(v) {\rm d}v$, where
$\zeta$ is another regular function. In fact, if we consider the
last ray $v=\bar v$ that passes through the center of the star
before the formation of the horizon, then to first order ${\rm d}V
= \zeta (\bar v){\rm d}v$, where $\zeta (\bar v)$ is now constant.

We can write near the horizon
\begin{equation}
{\rm d}U = \beta(0)\frac{{\rm d}{\cal U}}{{\rm d}u}{\rm d}u\;.
\end{equation}
Since for the center of the star ${\rm d}U = {\rm d}V = \zeta(\bar
v){\rm d}v$, this immediately integrates to
\begin{equation}
\zeta(\bar v)(v - {\bar v}) = \beta (0){\cal U}(u) =
         -\beta (0) \psi^{-1}\left(-u\right)
        \sim -2 \beta (0)\frac{M^2}{u}\;.
\end{equation}
The last approximation follows from Eq.\ (\ref{psi}) where
$\xi\sim \psi^{-1}(-2M^2/\xi)$ near the horizon.

Thus the late-time worldline for the center of the star is,
finally, represented by the equation\footnote{This result was also
obtained by Vanzo \cite{vanzo97} for a collapsing extremal thin
shell, but without considering a coordinate extension. Our method
is completely general and shows that Eq.\ (\ref{traj}) follows
only from the kinematics of collapse and the fact that the
external geometry is the extremal RN one.}
\begin{equation}
v \sim {\bar v} - \frac{A}{u}\;,\quad u\to +\infty\;, \label{traj}
\end{equation}
where $A = 2\beta (0)M^2/\zeta(\bar v)$ is a positive constant
that depends on the details of the internal metric and
consequently on the dynamics of collapse.

We first note that the worldline (\ref{traj}) differs from the one
resulting from the collapse of a nonextremal object, which would
be of the form (see e.g.\ \cite{FD77,Birrell82,saa})
\begin{equation}
v\sim {\bar v}-B{\rm e}^{-\kappa u}\;,\quad u\to +\infty\;.
\label{nextraj}\end{equation}
One immediately wonders, then, if our result can be recovered in
the case of nonextremal black holes by simply going to a higher
order approximation for the asymptotic worldline of the center of
the collapsing star. It is easy to see that this is not the case.
Recall that in the Kruskal coordinates ${\cal U}$ and ${\cal V}$,
the horizon is located at ${\cal U} = 0$. Say the worldline of the
center of the star crosses the horizon at some ${\cal V} =
\overline{\cal V}$. Let us expand ${\cal V({\cal U})}$ in a Taylor
series around ${\cal U} = 0$ such that ${\cal V} = \overline{\cal
V} +\alpha_1\,{\cal U}+\alpha_2\,{\cal U}^2$. The term
$\alpha_1\,{\cal U}\propto {\rm e}^{-\kappa u}$ is the usual one
found for the thermal case and $\alpha_2\,{\cal U}^2$ is the
correction. However, ${\cal U}^2 \propto {\rm e}^{-2\kappa u}$ and
so this term is also a constant for extremal incipient black
holes. In fact corrections are constant to arbitrary order. The
extremal worldline in no sense, therefore, represents a limit of
the nonextremal case but implies a real discontinuity in the
asymptotic behavior of the collapsing object.

Equations (\ref{traj}) and (\ref{nextraj}) contain the constants
$A$ and $B$, which are determined by the dynamics of collapse. In
the nonextremal case, it is known that no measurement performed at
late times can be used to infer the value of $B$, thus enforcing
the no-hair theorem. In particular, the spectrum of Hawking
radiation depends only on the surface gravity $\kappa$. It is
natural to ask whether a similar statement holds true also for
extremal black holes. This point will be analyzed in the following
sections.

\section{Bogoliubov coefficients}
\setcounter{equation}{0} \label{sec4}

Let us now consider a test quantum field in the spacetime of an
incipient extremal RN black hole. For the sake of simplicity, and
without loss of generality, we can restrict our analysis to the
case of a hermitian, massless scalar field $\phi$. Instead of
dealing with a black hole proper, we consider a two-dimensional
Minkowski spacetime with a timelike boundary
\cite{FD77,Birrell82}. This spacetime is described by null
coordinates $(u,v)$ and the equation governing the boundary is the
same that describes the worldline of the center of the star, say
$v=p(u)$.\footnote{In Refs.\ \cite{FD77,Birrell82,WF99} the
function $p$ is defined somewhat differently. For a generic shape
$x=z(t)$ of the boundary, one first defines a quantity $\tau_u$
through the implicit relation $\tau_{u}-z(\tau_{u})=u$. Then, the
function is defined as $p(u)=2\tau_{u}-u$, which is exactly the
phase of the outgoing component of the In modes, and $v=p(u)$ is
just the equation for boundary's worldline.} At the centre of the
star the ingoing modes of $\phi$ become outgoing, and vice versa;
this translates into the requirement that on the spacetime
boundary there is perfect reflection, or that $\phi(u,p(u))\equiv
0$. Hence, ``mirror": The timelike boundary in Minkowski spacetime
is traced out by a one-dimensional moving mirror for the field
$\phi$.

In general, for a worldline $v=p(u)$ one has ${\rm
d}\tau=\sqrt{p'(u)}\,{\rm d}u$, where $\tau$ is the proper time
along the worldline. From this and the fact that the acceleration
for the trajectory in two-dimensional Minkowski spacetime is
$a={1\over 2}\sqrt{p''(u)^2/ p'(u)^3}\ $, one can easily check
that Eqs. (\ref{traj}) and (\ref{nextraj}) yield $a^2=1/A$ and
$a^2=\kappa {\rm e}^{\kappa u}/(4B)$, respectively. Thus we see
that an incipient extremal black hole is modeled at late times by
a uniformly accelerated mirror; for nonextremal black holes the
acceleration of the mirror increases exponentially with time. In
both cases the mirror's worldline has a null asymptote $v=\bar{v}$
in the future, while it starts from the timelike past infinity
$i^-$ at $t=-\infty$.

Without loss of generality, one can assume that the mirror is
static for $t<0$. A suitable worldline is then
\begin{equation}
p(u)=u\Theta(-u)+f(u)\Theta(u)\;, \label{p(u)}
\end{equation}
where $\Theta$ is the step function, defined as
\begin{equation}
\Theta(\xi)=\left\{
\begin{array}{l}
1\quad\mbox{if}\quad\xi\geq 0\;,\\ 0\quad\mbox{if}\quad\xi <0\;,
\end{array}\right.
\end{equation}
and $f(u)$ is a function with the asymptotic form (\ref{traj}). In
order for the worldline to be $C^1$, $f(u)$ must be such that
$f(0)=0$ and $f'(0)=1$ . To simplify calculations, it is
convenient to choose $f(u)$ hyperbolic at all times
\cite{Birrell82}, i.e.,
\begin{equation}
f(u)=\sqrt{A}-{A\over u+\sqrt{A}}\;, \label{f(u)}
\end{equation}
which coincides with the function in the right hand side of Eq.
(\ref{traj}), up to a (physically irrelevant) translation of the
origin of coordinates.

Due to the motion of the mirror, one expects that the In and Out
vacuum states will differ, leading to particle production whose
spectrum depends on the function $p(u)$. In our case, because the
mirror worldline has a null asymptote $v=\bar{v}$ in the future
but no asymptotes in the past, the explicit forms of the relevant
In and Out modes for $\phi$ are easily shown to be
\begin{equation}
\phi_\omega^{\rm (in)}(u,v)={{\rm i}\over\sqrt{4\pi\omega}}
\left({\rm e}^{-{\rm i}\omega v}-{\rm e}^{-{\rm i}\omega
p(u)}\right) \label{in}
\end{equation}
and
\begin{equation}
\phi_\omega^{\rm (out)}(u,v)={{\rm i}\over\sqrt{4\pi\omega}}
\left({\rm e}^{-{\rm i}\omega u}- \Theta\left(\bar{v}-v\right){\rm
e}^{-{\rm i}\omega q(v)}\right)\;, \label{out}
\end{equation}
where $q(v)=p^{-1}(v)$ and $\omega>0$. The spectrum of particles
created in such a scenario is known, although, to our knowledge,
no one has pointed out the correspondence to the formation of
extremal black holes. However, since the result is something of a
textbook case, we here merely summarize the main steps; for
details, see e.g.\ Ref.\ \cite{Birrell82}, p\ 109.

The In and Out states of $\phi$ can be related by the Bogoliubov
coefficients:
\begin{equation}
\alpha_{\omega\omega'}=\left(\phi_\omega^{\rm (out)},
\phi_{\omega'}^{\rm (in)}\right)=-{\rm i}\int_0^{+\infty}{\rm
d}x\left[\phi_\omega^{\rm
(out)}(u,v)\rlpartial_t\phi_{\omega'}^{{\rm
(in)}}(u,v)^\ast\right]_{t=0}\;; \label{alphaa}
\end{equation}
\begin{equation}
\beta_{\omega\omega'}=-\left(\phi_\omega^{\rm (out)},
\phi_{\omega'}^{\rm (in)\ast}\right)= {\rm i}\int_0^{+\infty}{\rm
d}x\left[\phi_\omega^{\rm
(out)}(u,v)\rlpartial_t\phi_{\omega'}^{\rm
(in)}(u,v)\right]_{t=0}\;. \label{betaa}
\end{equation}
The spectrum of created particles is given by the expectation
value of the ``out quanta'' contained in the In state, $\langle
0,\mbox{in}|N_\omega^{\rm (out)}|0,\mbox{in}\rangle$. In terms of
the Bogoliubov coefficients this spectrum is
\begin{equation}
\langle N_\omega\rangle=\int_0^{+\infty}{\rm
d}\omega'\,|\beta_{\omega\omega'}|^2\;, \label{spectrum}
\end{equation}
where $\langle N_\omega\rangle$ is shorthand for $ \langle
0,\mbox{in}|N_\omega^{\rm (out)}|0,\mbox{in}\rangle$.

With the choice (\ref{f(u)}), one can compute Bogoliubov
coefficients that are appropriate in the asymptotic regime $t\to
+\infty$. Performing the integrals in Eqs.\ (\ref{alphaa}) and
(\ref{betaa}) gives \cite{Birrell82}
\begin{equation}
\alpha_{\omega\omega'}\approx {\rm i}\frac{\sqrt A}{\pi}{\rm
e}^{-{\rm i}{\sqrt A}(\omega+\omega')}
                K_1(2{\rm i}(A\omega\omega')^{1/2})\;,
\label{alpha}
\end{equation}
\begin{equation}
\beta_{\omega\omega'}\approx \frac{\sqrt A}{\pi}{\rm e}^{{\rm
i}{\sqrt A}(\omega-\omega')}
                K_1(2(A\omega\omega')^{1/2})\;,
\label{beta}
\end{equation}
where $K_1$ is a modified Bessel function, shown in Fig.\ 2. For
argument $z$, $K_1(z)\sim 1/z$ for $z\to 0$, and
$K_1(z)\sim\sqrt{\pi/(2z)}\,{\rm e}^{-z}$ when $z\to +\infty$
\cite{as}.

\begin{figure}[hbt]
\vbox{ \hfil \scalebox{0.40}{{\includegraphics{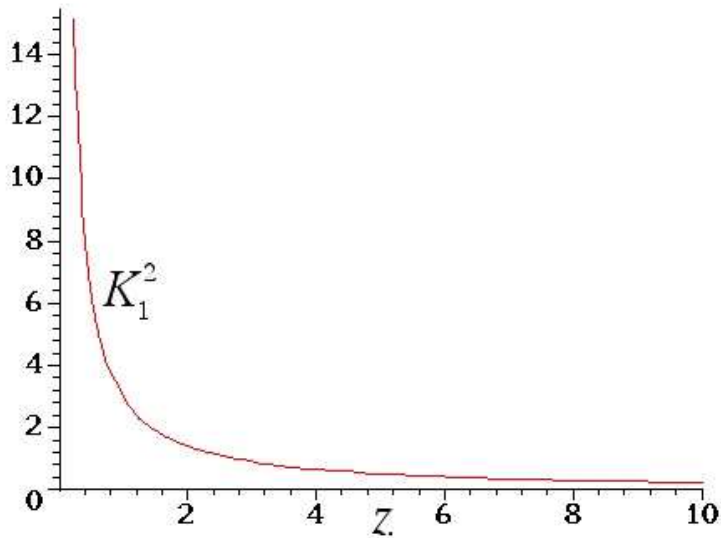}}}
\hfil }
\bigskip
\caption{
Plot of the modified Bessel function $K_1(z)$, squared.
} \label{bogol}
\end{figure}

We emphasize that Eqs.\ (\ref{alpha}) and (\ref{beta}) {\em do
not\/} correspond to a full evaluation of the integrals in Eqs.\
(\ref{alphaa}) and (\ref{betaa}), but only take into account the
contribution for $x\approx\sqrt{A}$, i.e., from the mirror
worldline at $u\to +\infty$. This is the only part of the
Bogoliubov coefficients that can be related to particle creation
by an incipient black hole, because any other contribution
corresponds to particles created much earlier, and depends
therefore on the arbitrary choice of $p(u)$ in the non-asymptotic
regime.\footnote{There has been some discussion in the literature
\cite{Grove86} about whether the calculation of the Bogoliubov
coefficients by Fulling and Davies \cite{FD77,Birrell82} is
correct. We find that their approximations are valid in the
asymptotic regime of interest to us.} Clearly, since $\langle
N_\omega\rangle\neq 0$, there is particle creation by the
incipient extremal RN black hole.\footnote{This result is only
apparently in contradiction with the analysis performed in Ref.\
\cite{vanzo97}, where it is claimed that there is no emission of
neutral scalar particles. In fact, such a conclusion was derived
for a massive field in the ultrarelativistic limit, and agrees
with the exponential behaviour of $K_1$ at large values of
$\omega$.}

Due to the $1/(\omega\omega')$ in the asymptotic form of
$|\beta_{\omega\omega'}|^2$, the spectrum (\ref{spectrum})
diverges at low frequencies. The divergence in $\omega'$ has the
same origin as the one that appears in the case of nonextremal
black holes, where
\begin{equation}
|\beta_{\omega\omega'}|^2 =\frac{1}{2\pi\omega'} \left(
\frac{1}{{\rm e}^{2\pi\omega/\kappa}-1} \right)\;. \label{thermal}
\end{equation}
These Bogoliubov coefficients also contain a logarithmic
divergence in $\omega'$, which is due to the evaluation of the
mode functions at $u=+\infty$ and for that reason can be
interpreted as an accumulation of an infinite number of particles
after an infinite time. The divergence can be removed, however, as
Hawking suggested \cite{Hawking75} by the use of wave packets
instead of plane-wave In states; this has the effect of
introducing a frequency cutoff.

Contrary to what happens in the nonextremal case, $\langle
N_\omega\rangle$ is not a Planckian distribution and therefore the
spectrum of created particles is nonthermal. Thus, the notion of
temperature is undefined. This result supports the view that an
extremal black hole is not the zero temperature limit of a
nonextremal one. However, it would be premature to base these
conclusions only on the basis of Eq.\ (\ref{spectrum}), because
the Bogoliubov coefficients tell us only that particles are
created {\em at some time\/} in the late stages of collapse, which
 does not necessarily mean that such creation takes place
at a steady rate. In the next two sections we refine our
conclusions through an analysis of the stress-energy tensor of the
quantum field.

\section{Preservation of cosmic censorship}
\setcounter{equation}{0} \label{sec5}

Equations (\ref{spectrum}) and (\ref{beta}) indicate that an
incipient extremal RN black hole creates particles with a spectrum
that depends on the constant $A$. These results immediately raise
two problems. First, since particle creation leads to black hole
evaporation, it seems that (some version of) the cosmic censorship
conjecture could be violated. Indeed, emission of neutral scalar
particles implies a decrease in $M$, while $Q$ remains constant;
evidently, a transition to a naked singularity ($Q^2>M^2$) should
take place. Second, the dependence of the spectrum on $A$, which
in turn depends on the details of collapse, raises the possibility
of getting information about the collapsing object through
measurements performed at late times, a contradiction of the
no-hair theorems.

We consider the first problem. The luminosity of the black hole,
the rate of change of $M$, is given by the flux of created
particles at infinity, or the $T_{uu}$ component of the
stress-energy tensor. Wu and Ford \cite{WF99} have recently
provided the expectation value of $T_{uu}$ for the case of a
moving boundary in two-dimensional Minkowski spacetime:
\begin{equation}
\langle :\!T_{uu}\!:\rangle =
\frac{1}{4\pi}\left(\frac{1}{4}\left(\frac{p''}{p'}\right)^2
        -{1\over 6}\frac{p'''}{p'}\right)\;.
\label{<:T:>}
\end{equation}
Inserting the form (\ref{p(u)}) of $p$, with $f$ given by Eq.\
(\ref{f(u)}), into Eq.\ (\ref{<:T:>}), one gets
\begin{equation}
\langle :\!T_{uu}\!:\rangle={1\over 24\pi\sqrt{A}}\, \delta(u)\;.
\end{equation}
Thus, the only nonvanishing contribution to $\langle
:\!T_{uu}\!:\rangle$ is due to the transition from uniform to
hyperbolic motion that takes place at $t=0$. For the discussion of
incipient black holes only the behaviour for $u\to +\infty$ is
relevant, and so this feature is uninteresting. On the other hand,
in the hyperbolic regime $\langle :\!T_{uu}\!:\rangle$ vanishes
identically. (It is also straightforward to check from Eq.\
(\ref{<:T:>}) that, conversely, a hyperbolic worldline is the only
one with nonzero acceleration that leads to $\langle
:\!T_{uu}\!:\rangle=0$.)

The result shows that the flux due to an incipient extremal black
hole vanishes asymptotically at late times. Consequently, extremal
black holes do not loose mass,\footnote{Here, we assume that
luminosity is simply related to $\langle:\!T_{uu}\!: \rangle$,
which amounts to assuming the validity of the semiclassical field
equation $G_{\mu\nu}=8\pi\langle :\!T_{\mu\nu}\!:\rangle$
\cite{DW}. This, however, might not be a good approximation when
$\phi$ is in a state with strong correlations (see, e.g.,
\cite{someth} and references therein).} and cosmic censorship is
preserved. However, the nonzero value of $\beta_{\omega\omega'}$
clearly shows that there {\em is\/} particle creation during
collapse. Cosmic censorship has apparently been rescued only at
the price of introducing a paradox, namely: Particles are created
{\em and\/} their flux has zero expectation value. How can these
two statements be simultaneously true?

This puzzling situation has been extensively discussed in the
context of particle emission from a uniformly accelerating mirror
\cite{FD77,Birrell82,FD76}. Fulling and Davies \cite{FD77} explain
the net zero energy flux in the presence of nonzero
$\beta_{\omega\omega'}$ by a special cancellation of the created
modes via quantum interference, which is due to contributions from
the coefficients $\alpha_{\omega\omega'}$. In the Appendix, we
analyze this issue further by examining the response function of a
detector.

\section{Preservation of the no-hair theorem}
\setcounter{equation}{0} \label{sec6}

We now turn to the second of the problems mentioned earlier: Given
that the spectrum contains the constant $A$, do extremal black
holes violate the no-hair theorems? The result $\langle
:\!T_{uu}\!:\rangle=0$ suggests an escape
--- in spite of the nonzero value of $\langle
N_\omega\rangle$, no radiation is actually detected. However, this
resolution raises new questions. If no radiation is detected, how
can one claim that the black hole emits anything at all? Is the
radiation observable? How should one then interpret $\langle
N_\omega\rangle$?

It is premature to claim that no radiation is detected only on the
basis of $\langle :\!T_{uu}\!:\rangle=0$, because there could be
other nonvanishing observables from which one might infer the
presence of quanta. A straightforward calculation shows that the
expectation values of $T_{vv}$ and $T_{uv}$ are also zero.
However, let us examine the variance $\Delta T_{uu}$ of the flux.
Wu and Ford \cite{WF99} have also recently given the following
expression for $\langle :\!T^{2}_{uu}\!:\rangle$ in the case of a
minimally coupled, massless scalar field in two-dimensional
Minkowski spacetime with a timelike boundary described by the
equation $v=p(u)$:
\begin{equation}
\langle :\!T^2_{uu}\!:\rangle= \frac{1}{\left(4\pi\right)^2}
\left(-\frac{4 p'^{2}}{(v-p(u))^4}+
\frac{3}{16}\left(\frac{p''}{p'}\right)^{4} -
\frac{1}{4}\frac{p'''}{p'}\left({p''\over p'}\right)^2+ {1\over
12}\left({p'''\over p'}\right)^2 \right)\;. \label{varT}
\end{equation}
If one ignores the so-called cross terms \cite{WF99}, this
coincides with the variance $\Delta T_{uu}$ (because in our case
$\langle :\!T_{uu}\!:\rangle=0$). With $p$ given by Eqs.\
(\ref{p(u)}) and (\ref{f(u)}), Eq.\ (\ref{varT}) gives, for $u>0$,
\begin{equation}
\langle :\!T^2_{uu}\!:\rangle=-{A^2\over 4\pi^2
\left(A+\left(v-\bar{v}\right)u\right)^4}\sim - {A^2\over
4\pi^2\left(v-\bar{v}\right)^4u^4}\;. \label{powlaw}
\end{equation}
Thus, in spite of the fact that the expectation value of the flux
vanishes identically, its statistical dispersion does not, but its
value becomes smaller and smaller and tends to zero in the limit
$u\to +\infty$. Hence, although one could in principle infer the
value of the constant $A$ by measuring the quantity $\Delta
T_{uu}$ at late times, such measurements will become more and more
difficult as $\Delta T_{uu}$ decreases according to Eq.\
(\ref{powlaw}). This damping is of course reminiscent of the
familiar damping of perturbations, which prevents one from
detecting by late-time measurements the details of an object that
collapses into a black hole \cite{price72}. And so, monitoring
$\Delta T_{uu}$ does not lead to a violation of the no-hair
theorem, because no trace of $A$ will survive in the limit $u\to
+\infty$.

This discussion shows only that no violation of the no-hair
theorem can be detected by measuring the variance in the energy
flux. The possibility remains that other types of measurement
could allow one to find out the value of $A$. If, however, $\Delta
T_{\mu\nu}\to 0$ for $u\to +\infty$, then the random variable
$T_{\mu\nu}$ must tend to its expectation value, i.e., to zero.
This means that, asymptotically, the properties of the field are
those of the vacuum state. Consequently, all local observables
will tend to their vacuum value.

Although extremal black holes obey the no-hair theorems, the {\em
way} in which cosmic baldness is enforced differs from the
nonextremal situation. Consider again the variance of the flux.
Inserting the function $p$ for nonextremal incipient black holes
(see Eq.\ (\ref{nextraj})) into Eqs.\ (\ref{<:T:>}) and
(\ref{varT}), one gets $\langle
:\!T_{uu}\!:\rangle=\kappa^2/(48\pi)$ and %
\begin{equation}
\langle :\!T_{uu}^2\!:\rangle={1\over (4\pi)^2}\left(
{\kappa^4\over 48}- {4\kappa^2B^2{\rm e}^{-2\kappa u}\over\left(
v-\bar{v}+B{\rm e}^{-\kappa u}\right)^4}\right) \sim
{\kappa^4\over 768\pi^2}- {\kappa^2B^2{\rm e}^{-2\kappa u}\over
4\pi^2\left(v-\bar{v}\right)^4}\;. \label{explaw}
\end{equation}
Contrary to the extremal case, the ``nonextremal" variance $\Delta
T_{uu}$ tends not to zero as $u\to +\infty$, but to the value
$\kappa^2\sqrt{2}/(48\pi)$, which corresponds to thermal emission;
this is sufficient to guarantee that no information about the
details of collapse is conveyed. Furthermore, the approach to this
value is exponentially fast, while for the extremal configuration
the decay obeys only a power law.

\section{Conclusions}
\setcounter{equation}{0} \label{sec7}

We have found a simple generalization of Kruskal coordinates that
allows us to examine the behavior of incipient, extremal RN black
holes. Although the coordinate transformation we employ is not
invertible in terms of elementary functions, it makes possible the
explicit calculation of the asymptotic form of the worldline for
the center of the collapsing object. Borrowing well-known results
from quantum field theory in the presence of moving boundaries, we
concluded that an incipient extremal black hole emits particles
with a nonthermal spectrum, which contains a constant that depends
on the details of gravitational collapse.

At first sight, this result seems to imply that the cosmic
censorship conjecture and the no-hair theorem are both violated by
extremal black holes. Closer scrutiny reveals that the flux of
emitted radiation vanishes identically, and in the limit $t\to
+\infty$ any measurement of local observables gives results
indistinguishable from those in the vacuum state. This is not
incompatible with a nonzero spectrum, which is not a local
quantity and tells us only that particles are created at some time
during collapse (not necessarily at $t=+\infty$). Thus, extremal
black holes are not pathological in this respect.

However, there are several clearly defined senses in which
nonextremal and extremal black holes differ. Information lost to
an external observer depends on the rate at which the statistical
dispersion of the flux approaches its value for $t\to +\infty$. In
the nonextremal case, the dispersion goes to zero exponentially
fast, Eq.\ (\ref{explaw}), whereas for an incipient extremal black
hole it follows a slower power law, given by Eq.\ (\ref{powlaw}).
More importantly, for $Q^2\to M^2$, Eq.\ (\ref{powlaw}) is not the
limit of Eq.\ (\ref{explaw})\footnote{Since $A$ and $B$ do not
depend on $u$ by definition, the only case that admits a
continuous limit is the one in which $A=0$. This cannot happen,
because it would correspond to a null worldline for the centre of
the star. Another apparent possibility, that $B\propto 1/\kappa$
so that $\kappa B$ is constant in the limit $\kappa\to 0$, is not
viable, because the right hand sides of Eqs.\ (\ref{powlaw}) and
(\ref{explaw}) would still have different functional dependences
on $u$. We thank Freeman Dyson for pointing out the issue to one
of us.}. One cannot therefore, consider quantum emission by an
incipient extremal black hole to be the limiting case of emission
by a nonextremal black hole. In particular, although at
$t=+\infty$ a black hole with $Q^2=M^2$ is totally quiescent it
would be incorrect to consider it as the thermodynamic limit of a
nonextremal black hole, that is, an object at zero temperature.
Indeed, the quantum radiation emitted by an incipient extremal
black hole is not characterized by a temperature at any time
during collapse. Whereas incipient nonextremal black holes have a
well defined thermodynamics, this is not true for extremal holes,
and they should be considered as belonging to a different class.
This result suggests that any calculations that implicitly rely on
a smooth limit in thermodynamic quantities at $Q^2 = M^2$ are
suspect, if not incorrect. Our conclusions, of course, are just
pertinent to incipient black holes; extending them to eternal
black holes seems plausible, but requires care. (Even at the
classical level, eternal black holes must be regarded as
fundamentally different from those deriving from collapse, because
the global structure of spacetime differs in the two cases.)

We close the paper drawing an analogy between the exotic subject
of particle production by extremal black holes and a well-known
piece of ordinary physics. The divergence of the particle spectrum
$\langle N_\omega\rangle$ is reminiscent of the infrared
catastrophe typical of QED, which manifests itself, for instance,
in the process of bremsstrahlung (see, e.g., Ref.\ \cite{mandl},
pp.\ 165--171). However, the infrared divergence in the
bremsstrahlung cross section produces no observable effect,
because it is canceled by analogous terms coming from radiative
corrections. (Thanks to the Bloch-Nordsieck theorem, this
cancellation is effective to all orders of perturbation theory.)
One may well wonder whether the $\omega=0$ singularity in our
spectrum is similarly fictitious and could thus be removed by
analogous techniques.

For the mirror this is possible, in principle, if one allows
momentum transfer from the field $\phi$ to the mirror, although
such a calculation is beyond the scope of the present paper. (See
Ref.\ \cite{par95} for a model that includes recoil.) But,
whatever the answer to the mirror problem might be, it does not
seem that one could transplant it in any straightforward way to
the case of an incipient black hole. Indeed, taking recoil into
account would amount to admitting that backreaction {\em is\/}
important and that the test-field approximation is never valid.
Thus, the whole subject would have to be reconsidered within an
entirely different framework.

In connection with the possible --- and crucial --- relevance of
backreaction, it is important to stress one aspect of our results.
We have seen in Sec.\ \ref{sec5} that if $Q^2=M^2$ at the onset of
the hyperbolic worldline (\ref{traj}), it will remain so and the
cosmic censorship conjecture is preserved. However, the fact that
mass loss from an incipient extremal black hole is zero
\emph{only} in the late hyperbolic stage seems to imply that an
enormous fine tuning is required in order to produce an extremal
object by means of gravitational collapse. In fact, an object that
is extremal from the start of its collapse might be unstable with
respect to the transition to a configuration with $Q^2>M^2$. Such
a transition would be triggered by quantum emission in the early
phases of collapse, when $p(u)$ has not yet assumed its hyperbolic
form. This raises the question of how, in presence of quantum
radiation, the formation of a naked singularity is prevented
(e.g., by the emission of charged particles) and the cosmic
censorship conjecture preserved.\\


\noindent{\bf Note added} 
Simultaneously with this work Anderson, Hiscock and
Taylor~\cite{AHT00} have demonstrated that for static RN geometries,
zero-temperature black holes cannot exist if one considers spacetime
perturbations due to the back reaction and quantum fields.


\noindent{\bf Acknowledgements} It is a pleasure to thank J.\
Almergren for support in drawing Fig.\ 1 and F.\ Belgiorno for helpful
remarks on a first draft of the paper. S.L. thanks R.\ Parentani for
stimulating discussions.  T.R.\ would like to acknowledge
the hospitality of S.I.S.S.A., where this work was carried out.

\section*{Appendix: Detecting radiation from a uniformly
accelerated mirror} \setcounter{equation}{0}
\renewcommand{\theequation}{A.\arabic{equation}}

In Sec.\ \ref{sec5} we mentioned the apparently paradoxical
situation in which nonzero particle production (as shown by
nonvanishing Bogoliubov coefficients $\beta_{\omega\omega'}$) is
accompanied by zero energy flux (vanishing expectation value of
the stress-energy-momentum tensor.) Discussions about such issues
are often phrased in terms of ideal detectors
\cite{Birrell82,DeWitt79}. Although our arguments in the body of
the paper are based solely on the behavior of the
stress-energy-momentum tensor, we can gain some additional insight
into the ``paradox'' by considering the response of a simple
monopole detector on a geodesic worldline $v=u+2x_0$, with
$x_0=\mbox{const}$, in two-dimensional Minkowski spacetime.

We are interested in computing the detector response function per
unit time, defined as
\begin{equation}
{\cal R}(E)=\lim_{T\to +\infty} {1\over 2T}\int_{-T}^{T}{\rm
d}\tau\int_{-T}^{T} {\rm d}\tau'\, \Theta(E){\rm e}^{-{\rm
i}E(\tau - \tau')}D^+(u(\tau),v(\tau);u(\tau'),v(\tau'))\;,
\label{response}
\end{equation}
where $D^+$ is the Wightman function of the scalar field in the In
vacuum, $u(\tau)=\tau-x_0$, $v(\tau)=\tau+x_0$, and $E$ is the
excitation energy of the detector. (Note that $E\geq 0$, which is
automatically enforced by the presence of the step function
$\Theta(E)$ in the right hand side of Eq.\ (\ref{response}).) In
terms of the In modes, $D^+$ has the form
\begin{equation}
D^+(u,v;u',v')=\int_{-\infty}^{+\infty}{\rm d}\omega\,
\Theta(\omega)\phi_\omega^{\rm (in)}(u,v) \phi_\omega^{\rm
(in)}(u',v')^\ast\;, \label{orco}
\end{equation}
where we have extended the integration range to $-\infty$, by
introducing the step function $\Theta(\omega)$.

Since the definition of ${\cal R}(E)$ involves an integration over
time from $-\infty$ to $+\infty$, in the case of a mirror
worldline of the type (\ref{p(u)}) it will get contributions
corresponding to the nonzero flux (like, e.g., the one at $u=0$
when $f(u)$ is given by Eq.\ (\ref{f(u)})). These we regard as
spurious, because we are really interested in clarifying the
relationship between zero flux and nonzero spectrum in the
hyperbolic regime. For this reason, let us consider a mirror
worldline which is hyperbolic at all times, say $p(u)=-A/u$ for
$u>0$, for which there can be no such spurious contributions to
${\cal R}(E)$.

The worldline $p(u)=-A/u$ has a null asymptote in the past, thus
\begin{equation} \phi_\omega^{\rm (in)}(u,v)={{\rm
i}\over\sqrt{4\pi\omega}} \left({\rm e}^{-{\rm i}\omega v}-
\Theta(u){\rm e}^{-{\rm i}\omega p(u)}\right)\;. \label{mode}
\end{equation}
On substituting Eq.\ (\ref{mode}) into Eq.\ (\ref{orco}), we have
\begin{equation}
D^+(u,v;u',v')=F_1(v,v')+F_2(u,v')+F_3(v,u')+F_4(u,u')\;,
\end{equation}
where:
\begin{equation}
F_1(v,v')={1\over 4\pi}\int_{-\infty}^{+\infty}{\rm d}\omega\,
{\Theta(\omega)\over |\omega|}\,{\rm e}^{-{\rm i}\omega (v-v')}\;;
\end{equation}
\begin{equation}
F_2(u,v')=-{1\over 4\pi} \Theta(u)\int_{-\infty}^{+\infty}{\rm d}
\omega\,{\Theta(\omega)\over |\omega|}\, {\rm e}^{{\rm i}\omega
(v'-p(u))}\;;
\end{equation}
\begin{equation}
F_3(v,u')=-{1\over 4\pi} \Theta(u')\int_{-\infty}^{+\infty}{\rm d}
\omega\,{\Theta(\omega)\over |\omega|}\, {\rm e}^{-{\rm i}\omega
(v-p(u'))}\;;
\end{equation}
\begin{equation}
F_4(u,u')={1\over 4\pi}\Theta(u)\Theta(u')\int_{-\infty}^{+\infty}
{\rm d}\omega\,{\Theta(\omega)\over |\omega|}\, {\rm e}^{-{\rm
i}\omega (p(u)-p(u'))}\;.
\end{equation}
Correspondingly, ${\cal R}(E)$ can be split into four parts:
${\cal R}(E)={\cal R}_1(E)+{\cal R}_2(E)+{\cal R}_3(E)+{\cal
R}_4(E)$.

The terms ${\cal R}_1(E)$, ${\cal R}_2(E)$, and ${\cal R}_3(E)$
can be computed straightforwardly, by using the formal identities
\begin{equation}
\lim_{T\to +\infty}\int_{-T}^T{\rm d}\tau\,{\rm e}^{{\rm
i}\xi\tau}=2\pi\delta(\xi)
\end{equation}
and
\begin{equation}
{1\over |E|}\,\Theta(E)\Theta(-E)=2\delta(E)\;, \label{gnarf}
\end{equation}
the latter being easily established by considering the sequence of
functions $(| E | +\epsilon)^{-1}
\Theta(E+\epsilon)\Theta(-E+\epsilon)$ in the limit $\epsilon\to
0$. We get ${\cal R}_1(E)=-2{\cal R}_2(E)=-2{\cal R}_3(E)=
\delta(E)$, so the first three contributions to ${\cal R}(E)$ sum
to zero.

The computation of ${\cal R}_4(E)$ is cleaner if one works in
dimensionless variables, such as $\widetilde{E}=\sqrt{A}\,E$,
$\tilde{\omega}=\sqrt{A}\,\omega$, $\tilde{\tau}=\tau/\sqrt{A}$.
The identity
\begin{equation}
\int_0^{+\infty}{\rm d}\tilde{\omega}\,{{\rm e}^{-{\rm
i}\tilde{\omega} (\xi-{\rm
i}0)}\over\tilde{\omega}}=-\ln\left(\xi-{\rm i}0\right)+I-{\rm
i}\,{\pi\over 2}\;,
\end{equation}
where $I$ is the divergent quantity
\begin{equation}
I=\int_0^{+\infty}{\rm d} \tilde{\omega} \, {\cos\tilde{\omega}
\over \tilde{\omega}}\;,
\end{equation}
together with the properties of the logarithm, allows us to write
\begin{eqnarray}
\int_{-\infty}^{+\infty}{\rm d}\tilde{\omega}\,
{\Theta(\tilde{\omega})\over |\tilde{\omega}|}\, \exp\left(-{\rm
i}\tilde{\omega}{\tilde{\tau}-\tilde{\tau}'
\over\left(\tilde{\tau}-\tilde{x}_0\right)\left(\tilde{\tau}'
-\tilde{x}_0\right)}\right)=\int_{-\infty}^{+\infty}{\rm d}
\tilde{\omega}\,{\Theta(\tilde{\omega})\over |\tilde{\omega}|}\,
{\rm e}^{-{\rm i}\tilde{\omega}\left(\tilde{\tau}
-\tilde{\tau}'\right)}\nonumber\\ -\int_{-\infty}^{+\infty} {\rm
d}\tilde{\omega}\, {\Theta(\tilde{\omega})\over
|\tilde{\omega}|}\, {\rm e}^{-{\rm
i}\tilde{\omega}\left(\tilde{\tau}-
\tilde{x}_0\right)}-\int_{-\infty}^{+\infty}{\rm d}
\tilde{\omega}\,{\Theta(\tilde{\omega})\over |\tilde{\omega}|}\,
{\rm e}^{{\rm i}\tilde{\omega}\left(\tilde{\tau}'-
\tilde{x}_0\right)}+2I\;.
\end{eqnarray}
In this expression we have replaced one of the quantities $I-{\rm
i}\pi/2$ with its complex conjugate by simultaneously changing the
sign in one of the exponents. This manipulation is allowed by the
fact that, since $\tilde{\tau}-\tilde{x}_0$ and
$\tilde{\tau}'-\tilde{x}_0$ can never become negative in $F_4$,
their logarithms are always real. Note that the resulting
expression agrees with the property of ${\cal R}(E)$ of being a
real quantity.

We can thus write ${\cal R}_4(E)={\cal R}_{41}(E)+{\cal
R}_{42}(E)+{\cal R}_{43}(E)+{\cal R}_{44}(E)$. Using the formal
relations
\begin{equation}
\lim_{\widetilde{T}\to +\infty}
\int_{\tilde{x}_0}^{\widetilde{T}}{\rm d}\tilde{\tau}\, {\rm
e}^{{\rm i}\xi\tilde{\tau} }=\pi\delta(\xi)+{\rm i} {\rm e}^{ {\rm
i}\,\xi\tilde{x}_{0} }\,{\rm P}\left(1 \over\xi\right)\;,
\end{equation}
and
\begin{equation}
\lim_{\widetilde{T}\to +\infty}{1\over\widetilde{T}}
\int_{\tilde{x}_0}^{\widetilde{T}}{\rm d}\tilde{\tau}\, {\rm
e}^{{\rm i}\xi\tilde{\tau}}=\Theta(\xi)\Theta(-\xi)\;,
\end{equation}
together with Eq.\ (\ref{gnarf}), we get ${\cal
R}_{41}(E)=\delta(E)/4$, ${\cal R}_{42}(E)+{\cal
R}_{43}(E)=-\delta(E)/2$, and ${\cal R}_{44}(E)=I\delta(E)/4$.
Finally, since $I$ is divergent we can write
\begin{equation}
{\cal R}(E)={I\over 4}\,\delta(E)\;. \label{F}\end{equation}

Thus, we have essentially a delta function peaked at zero energy.
Now, ${\cal R}(E)$ is related to a quantum mechanical probability
and so this result means that, for any value $E>0$ of the energy,
no matter how small, the detector has probability 1 of making a
transition of amplitude smaller that $E$ and probability 0 of
detecting particles of higher energy. (Of course, this does not
mean it will {\em never\/} make transitions with $E>0$; only,
these take place with probability 0.) The reason for this
behaviour is evidently the divergence in the spectrum as
$\omega\to 0$. Of course, the detector does not gain energy during
such a ``detection'' --- in fact, one can say that there is no
detection at all. This is compatible with the zero value of the
flux.

Hence one might be tempted to call the particles emitted by a
mirror in hyperbolic motion ``phantom radiation'': Because only
arbitrarily soft particles would be registered by the detector
with any nonzero probability, there would be no chance to
determine the spectrum $\langle N_\omega\rangle$. The question
arises therefore, whether there is any way to screen our detector
from this overwhelming flux of soft quanta.

One might think to act on the selectivity of the detector by using
a two level system that requires at least a minimal energy to
switch. Unfortunately, the detection of the infinite tail of soft
quanta corresponds to the ``transition'' from the ground state to
the ground state, and there is obviously no way to forbid this
process. The detector cannot be forbidden to not switch!

So also the analysis of the response function of a detector seems
to prove that the radiation from uniformly accelerated mirrors
(and extremal incipient black holes) is in some sense like the
apple in Dante's purgatory: We can see it with our mind but we
shall never have it in our hands...


\end{document}